# Towards a Global Scale Quantum Information Network: A Study Applied to Satellite-Enabled Distributed Quantum Computing

Laurent de Forges de Parny[1], Luca Paccard[1], Mathieu Bertrand[1], Luca Lazzarini[1], Valentin Leloup[1], Raphael Aymeric[1], Agathe Blaise[2], Stéphanie Molin[2], Pierre Besancenot[1], Cyrille Laborde[1], and Mathias van den Bossche[1]

[1]*Thales Alenia Space, 26, Avenue J-F Champollion, 31037, Toulouse, France*
[2]*Thales SIX GTS, 4 Avenue des Louvresses, 92230 Gennevilliers, France*

**ABSTRACT**

Recent developments have reported on the feasibility of interconnecting small quantum registers in a quantum information network of a few meter-scale for distributed quantum computing purposes. This multiple small-scale quantum processors communicating and cooperating to execute computational tasks is considered as a promising solution to the scalability problem of reaching more than thousands of noise-free qubits. Here, we propose and assess a satellite-enabled distributed quantum computing system at the French national scale, based on existing infrastructures in Paris and Nice. We consider a system composed of both a ground and a space segment, allowing for the distribution of end-to-end entanglement between Alice in Paris and Bob in Nice, each owning a few-qubit processor composed of trapped ions. In the context of quantum computing, this entanglement resource can be used for the teleportation of a qubit state or for gate teleportation. We numerically assess the entanglement distribution rate and fidelity generated by this space-based quantum information network, and discuss concrete use cases and service performance levels in the framework of distributed quantum computing.

## 1. INTRODUCTION

Quantum entanglement is the basic resource to be managed in Quantum Information Networks (QINs), enabling quantum state teleportation between two distant end-users. The functionalities of a QIN aim at producing, transmitting, and exploiting entanglement. The role of a QIN is therefore to deliver to the end users, hereafter called Alice and Bob, a sufficient amount of quantum entanglement in a well-defined state, typically one of the four Bell states, with a sufficient fidelity for allowing quantum communication use cases (e.g., Quantum Key Distribution, Quantum teleportation, etc.). A promising perspective of QIN in quantum computing is their use as a connector between several quantum chipsets, each with a small number of physical qubits. Networking of quantum chipsets allows, in principle, the exponentiation of the computational resource. Furthermore, the use of quantum chipsets with a small number of qubits in a network has the advantage of easing the management of the error correction on each chipset and therefore reduces the overall source of noise [1]. This modular networking approach can offer a solution to the scalability problem of developing a quantum chipset with a very large number of qubits required for solving customer problems. This scaling-up approach is already investigated by XANADU with photonic links using 35 photonic chips in a network to demonstrate the functionality and feasibility of a (sub-performant) scale model of a 12-physical qubit mode quantum computer, called AURORA [2]. Modular quantum-computing is also central to IBM's quantum roadmap [3]. IBM has designed a tunable-coupler quantum processor called Flamingo, which pairs two 156-qubit Heron processors with a built-in quantum communication link. IBM plans to connect up to seven Herons processors to create a modular Flamingo processor with more than 1,000 qubits, including quantum transduction [4]. Very recently, a very basic version of Grover's search algorithm running across two interconnected processors had been demonstrated at Oxford University by using a photonic connection between two trapped-ion qubits [5]. The aforementioned examples consider local connections between the qubits, typically of the order of a few meters in the laboratory.

However, the need for connections between remote qubits for a global quantum Internet is already under active study all around the world in research labs and in industry [6][7]. Among others, we can cite the network development in the USA, Spain, China, and France. In the USA, quantum-enabling networking architectures have been implemented over 158 km [8], with entanglement distribution across 140 km of optical fiber [9], and over 34 km of deployed fiber with high-entangled rates and fidelity in automated mode [10]. Quantum memories are also integrated in these developments [11]. In Spain, similar demonstrations of entanglement distribution in fiber-based networks with integrated quantum memories are under study, with a different platforms [12][13]. China is also developing a metropolitan-scale testbed for the evaluation and exploration of multi-node quantum network protocols for the quantum Internet [14]. In France, the realization of an operational entanglement-based three node metropolitan quantum network over a total distance of 50 km has been reported [15] and commercial quantum memories are also under development [16]. All these examples clearly show the need to increasing the range of the network, while being affected by large losses in the optical fibers. As demonstrated by China in 2017, the low Earth orbit satellite offers a competitive solution for long-distance entanglement distribution compared to ground optical fibers [17], despite a promising strategy of range extension under development in ground fiber network with entanglement swapping [18][19].

In this article, we investigate the level of performance of satellite-enabled distributed quantum computing at the French national scale between Paris and Nice. Alice in Paris and Bob in Nice receive quantum entanglement from a QIN made of Local Area Quantum Networks (LAQN) in Paris and Nice connected with a Low Earth Orbit (LEO) satellite embarking an entangled photons source. This end-to-end entanglement provides a physical resource allowing Alice and Bob to connect their respective small-scale quantum chipset for implementing non-local unitary coupling gates, similarly to the recent Oxford experiment at lab scale [5]. We consider here a single satellite passage in the case of clear sky (i.e. no clouds). The goal of this paper is therefore to emphasize the role of a space-based QIN in distributing quantum entanglement for a practical distributed quantum computing application, hereafter the teleportation of a controlled-Z gate. The article is structured as follows: Section 2 describes the generic QIN network architectures as a function of the distance between the end-users in order to introduce the elements of the network and the satellite requirements for long-distance entanglement distribution. Section 3 presents the architecture implementation of the space-based QIN studied in this article and the use case under consideration in terms of distributed quantum computing. Section 4 presents the QIN performance in terms of entanglement and fidelity distributed by the QIN and the number of gates teleportation between Alice and Bob. Section 5 summarizes the discussions and provides perspectives.

## 2. GENERIC QIN NETWORK ARCHITECTURES

Distributed or blind quantum computing requires the sharing of quantum entanglement distributed by the QIN. Depending on the distance between Alice and Bob, many architectures are foreseen (Figure 2-1). For a short distance (typically < 100 km), a single entangled photon source placed between Alice and Bob could be sufficient for distributing entanglement between them. For longer distances, the optical fiber losses (~0.2 dB/km at best) compromises this solution. A tricky solution considered in the literature [18], and used in different research laboratories, is to plug in a chain of entangled photon sources and Bell state measurement (BSM) with quantum memories as entanglement swapper (improperly named "quantum repeater"), see Figure 2-1 b). Each entangled photon source emits pairs of photons towards a BSM device with the strategy "repeat until success" of the entanglement swapping. The role of the BSM is to measure the correlations between a pair of photons that have never interacted before, resulting in an entanglement between the remaining pair of photons. This allows to effectively extend the range of the entanglement over the distance. The use of multimodal quantum memories located at the BSM devices is necessary for the repeater architecture to provide an advantage. Quantum memories allow for the success probability of entanglement distribution to scale with the distance of a single repeater link and the number of BSM devices whereas without quantum memories this success probability scales with the total distance between Alice and Bob. However, since a BSM allows for the distinction of only two of the four Bell states (at least with linear optics), the maximal theoretical BSM efficiency is ½. Therefore, the efficiency of this solution decreases with the number of BSM devices, i.e. with the distance between Alice and Bob. For very long distances (e.g., > 500 km), the satellite could offer a better efficiency for distributing entanglement to the end-users (Figure 2-1 c). Particularly, a satellite at low Earth orbit with large telescopes is preferred, as demonstrated by the Chinese mission with the Micius satellite [17]. Note that the QIN is embedded in a classical communication network for all architectures of Figure 2-1. The classical communication network is required for the management/coordination of the equipment and for the unitary correction in the entanglement swapping protocol [21].

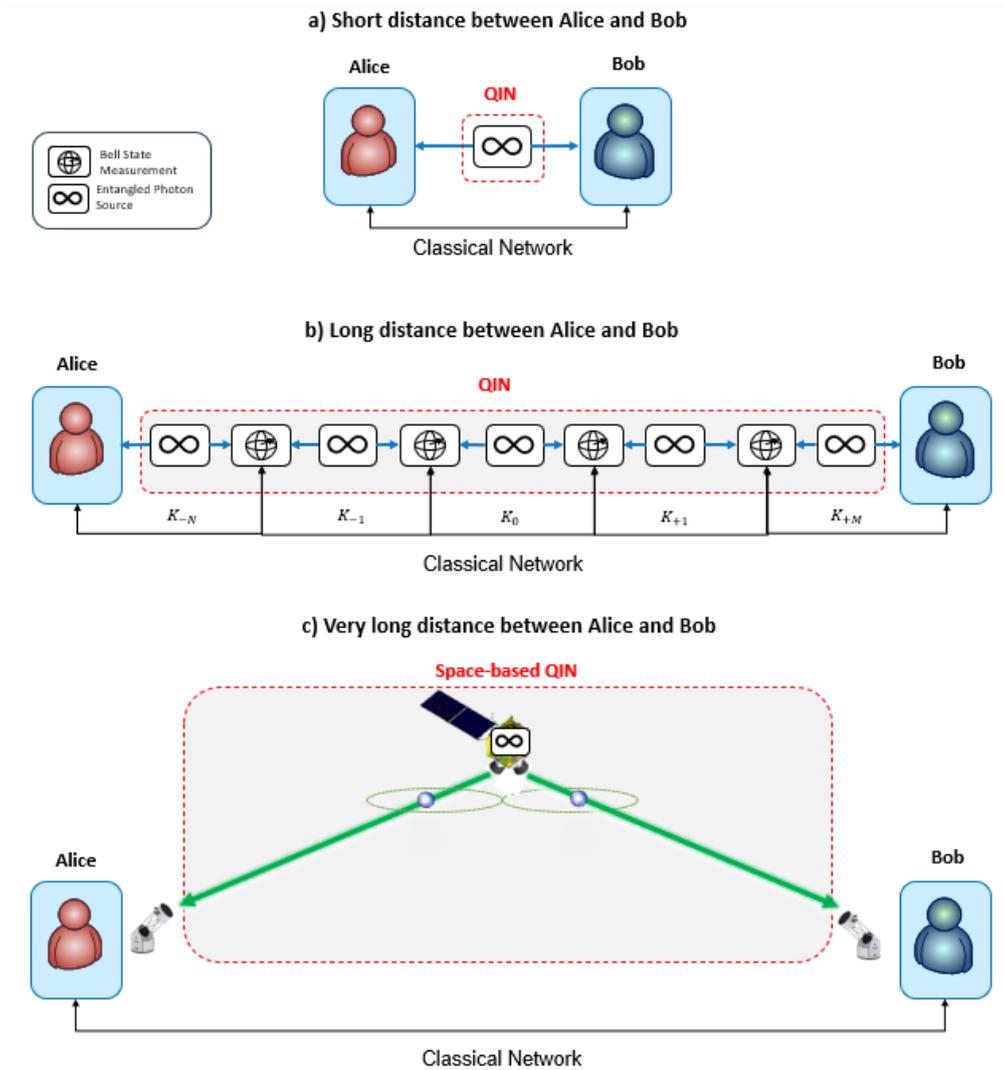

Figure 2-1: Architectures of QIN for a) short, b) long, and c) very long distances between the users, Alice and Bob.

## 3. PROPOSAL OF IN FIELD IMPLEMENTATION OF A SPACE-BASED QIN NETWORK

In this section, we propose a space-based QIN architecture with realistic constraints concerning the implementation. We suppose Alice is located in Paris and Bob is located in Nice at roughly 700 km away from Alice. The distance requires the use of a satellite in the QIN for distributing entanglement to Alice and Bob. The QIN of interest is composed of one LEO satellite and two Local Area Quantum Networks (LAQN) in Paris [20][22] and in Nice Côte d'Azur [23][24]. In the proposed implementation, Alice and Bob are not directly connected to a quantum ground station, thus requiring the use of both a space and two ground segments for entanglement distribution. Indeed, quantum ground station with telescope's aperture larger than 1 m are not flexible for integration close to the user, unfortunately [25]. The architecture is therefore a mixture between the architectures depicted in Figure 2-1 b) and c). The network design also takes into account recent developments in France, in Paris and in Nice. Although BSM and quantum memories have not been yet implemented in this network, some nodes and optical ground stations already exist as explained hereafter.

The LAQN in Paris comprises three nodes:
- Alice: a quantum chip, with one BSM and one quantum memory, located at LIP6 Sorbonne University Laboratory;
- An entangled photon source at ORANGE LAB at Châtillon;
- A BSM device, with two quantum memories and with an optical ground station (i.e., a telescope), at THALES Palaiseau.

The LAQN in Nice Côte d'Azur also comprises three nodes:
- Bob: a quantum chip, with one BSM and one quantum memory, located at INPHYNI Côte d'Azur University Laboratory;
- An entangled photon source at INRIA Sophia Antipolis;
- A BSM device, with two quantum memories and with an optical ground station (MéO [25]), at Calern.

Connecting those two local areas, separated by roughly 700 km, with a single optical fiber link will introduce more than 140 dB loss - assuming 0.2 dB/km losses at best - and will require the use of many BSM nodes on the ground for entanglement swapping. Propagating entanglement with ground links over such distances will be highly inefficient and is not yet allowed with state-of-the-art technologies. However, using a LEO satellite for connecting two nodes at such distances offers much better performance, as already shown by the Chinese demonstration with the Micius satellite [17]. We therefore propose to use a LEO satellite at 600 km altitude for connecting the Paris and Nice LAQNs (see Ref [7] for system trade-offs). This network scheme is depicted in Figure 3-1.

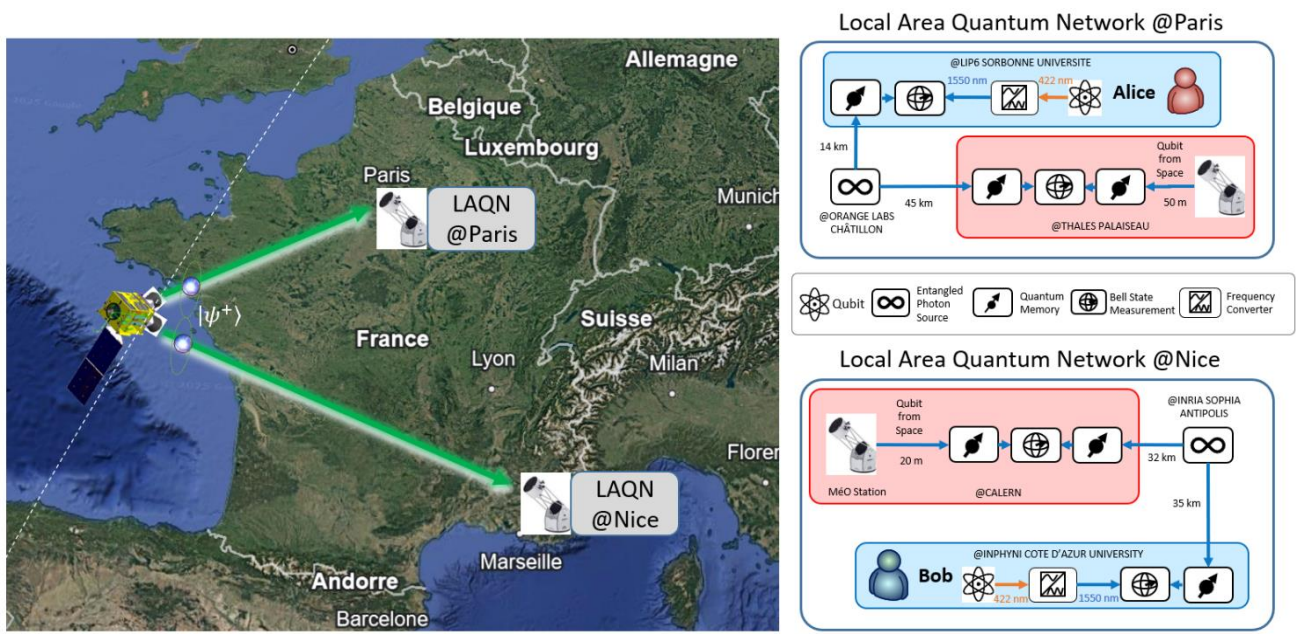

Figure 3-1: Scheme of the proposed space-based QIN composed of one LEO satellite and two local area quantum networks in Paris and Nice.

For the sake of simplicity, all the sources (Alice and Bob, and the entangled photon sources), quantum memories, and BSM are supposed to have the same performance, see the parameters in Table 1 in the Annex. All the sources of the quantum network have the same rate $R_{src}$ for the sake of simplicity. The entangled photon sources are supposed to be based on SPDC nonlinear process with quasi-degenerate idler and signal at 1550 nm in polarization encoding. The BSM are using state-of-the-art Superconducting Nanowire Single Photon Detectors (SNSPD) with ideal theoretical BSM efficiency, i.e., $\eta_{BSM} = 0.5$. In our model, we assume multimodal quantum memories with very optimistic quantum memory fidelity and time storage (see Table 1). Also, we suppose having writing announcement (heralding) function. These assumptions on quantum memories are by far the strongest hypothesis of this present study. Furthermore, we suppose the BSM devices embedded into a classical communication network (e.g., Internet) for the required classical communication in the quantum teleportation protocol (Bell state measurement announcement and associated unitary operation correction [21]) and BSM management.

**Entanglement distribution model**:

We have developed a model for the number of the received entangled photons pairs at the end-user nodes and a model for the associated fidelity. The particularity of our models is that we consider multimodal quantum memories for increasing the success probability of the entanglement swaps.

The rate of the end-to-end received entangled photon pairs to Alice and Bob is given by:

$$\sigma_{pair}^{end-to-end} = \frac{R_{src}}{N} \left[ \eta^{Alice} \cdot \eta_{swap} \cdot \eta_{elem}^{Paris} \cdot \eta_{swap} \cdot \eta_{elem}^{Sat} \cdot \eta_{swap} \cdot \eta_{elem}^{Nice} \cdot \eta_{swap} \cdot \eta^{Bob} \right], \quad (1)$$

with:
- $\frac{R_{src}}{N}$ the swapping rate, given by the source rate $R_{src}$ divided by the number of modes $N$ of the quantum memories (the swaps are performed after $N$ time steps, see Entanglement swapping procedure discussion below);
- $\eta^{Alice} = \eta^{Bob} = R_{src} \cdot \eta_{conv} \cdot \eta_{QM}$, with $\eta_{conv}$ the wavelength conversion efficiency from 422 nm to 1550 nm (see use case section) and $\eta_{QM}$ the efficiency of quantum memory, given by Eq. (7) of the Annex paragraph;
- $\eta_{swap} = \eta_{BSM} \cdot \eta_{det}^2$ with $\eta_{BSM}$ the efficiency of the Bell state measurement device and $\eta_{det}$ the single photon SNSPD detector efficiency;
- $\eta_{elem}^{Paris}$ and $\eta_{elem}^{Nice}$ the efficiency of an elementary ground path, given by Eq. (10) of the Annex paragraph;
- $\eta_{elem}^{Sat}$ the efficiency of the elementary space path, given by Eq. (11) of the Annex paragraph.

The fidelity of the end-to-end received entangled photon pairs, based on the Werner states, is given by [27][28][29]:

$$F_{end-to-end} = \frac{1 + 3\mathcal{W}_{end-to-end}}{4} \quad (2)$$

with $\mathcal{W}_{end-to-end}$ the Werner parameter of the end-to-end path given by Eq. (20) in the Annex. For the space distribution of entangled photons, we take into account the stray photons mixing with qubits from the satellite. However, the field of view used in our model takes into account single mode fiber coupling at the ground level, which results in a small percentage of stray photons entering into the quantum memories. Other elements degrading the fidelity are the source and quantum memory imperfections, as well as the free space channel imperfection (misalignment). Concerning the ground path, the fidelity degradation is due to the quantum repeater imperfections, as well as the impairments in the optical fibers.

The list of all the parameters used in our simulations is provided Table 1 in the Annex. The parameters used for the quantum memories are significantly optimistic but seems a good target for an operational QIN. The orbital simulations are performed with ANSYS Systems Tool Kit®, a powerful commercial software enabling a mission scenario description including a detailed and realistic orbital propagation, taking into account, e.g., the non-sphericity of Earth, the residual atmosphere drag on the satellite as a function of its weight, geometry, and orientation (attitude sequences), as well as the ground station locations. We do not consider entanglement purification in this study.

**Entanglement swapping procedure**:

Our time-iterative simulations are performed with a discrete time step $\delta t = 1/R_{src}$, with $R_{src}$ the entangled photon pair source rate. At each time step, the entangled photon pairs sources can send an entangled photon pair on each elementary links of the network. We voluntarily choose a sub-optimal procedure - associated to the lower bound for the rate of end-to-end received entangled photon pairs - for the entanglement swapping assuming the Bell state measurements are synchronized after the timeslot duration of $\Delta t = N \cdot \delta t$, with $N$ the number of memory storage modes. While an optimal procedure would require an optimization of the Bell state management considering the successful storage on each elementary links over each time step, our procedure gives a rough idea of the lower non-trivial performance of the whole chain. In other words, after time $\Delta t$, the end-to-end (Alice-Bob) entanglement can be successfully established or not according to our hypotheses.

**Quantum distributed computing use case**:

We propose to focus on the distributed quantum computing use case of Ref [5] where the entangled photon pairs received at the end points of the QIN, i.e., at Alice and Bob nodes, will be used for non-local two-qubit gates teleportation. The simplest use case is the deterministic teleportation of a non-local controlled-Z gate requiring only two qubits on each quantum chipset: one network qubit and one circuit qubit. Such a gate teleportation consumes, in principle, only one Bell pair and the exchange of two classical bits. It is therefore very practical to focus on the teleportation of a controlled-Z gate as a key performance indicator, while we should keep in mind that any arbitrary two-qubit unitary operation can be decomposed into at most three controlled-Z gates, with three instances of quantum gate teleportation [5]. An example of a physical platform compatible with QINs is trapped ions, that could be plugged with a photonic QIN without transducers, as shown in Ref [5] where $^{88}$Sr$^+$ ($^{43}$Ca$^+$) ion is used as a network (circuit) qubit. However, a wavelength conversion, with efficiency $\eta_{conv}$, is necessary at the output of the quantum chipset for conversion from 422 nm to 1550 nm. The trapped ion being emissive, both Alice and Bob need a local BSM for interfacing with the space-based QIN.

As depicted in Figure **3-1**, the space-based QIN provides entanglement to Alice and Bob's network nodes in state $|\psi^+\rangle_{23} = (|H\rangle_2|V\rangle_3 + |V\rangle_2|H\rangle_3)/\sqrt{2}$, in agreement with Ref [5]. Figure **3-1** can be simplified as shown in Figure 3-2 where the entangled photon pair in state $|\psi^+\rangle_{23}$ is sent to Alice and Bob nodes.

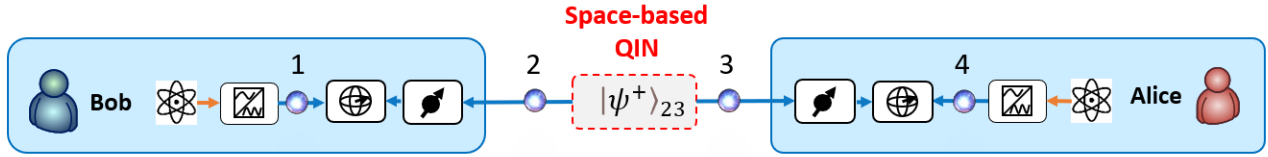

Figure 3-2: Simplification of Figure 3-1 where the space-based QIN is represented by the red dotted box.

We suppose Alice and Bob qubit in a hybrid spin-photon entangled state of the form $|\psi\rangle_1 = (|H\rangle_1|\downarrow\rangle_1 + |V\rangle_1|\uparrow\rangle_1)/\sqrt{2}$ and $|\psi\rangle_4 = (|H\rangle_4|\downarrow\rangle_4 + |V\rangle_4|\uparrow\rangle_4)/\sqrt{2}$, respectively.

The four-state photon is given by:

$$|\psi\rangle_1 \otimes |\psi^+\rangle_{23} \otimes |\psi\rangle_4 = \frac{1}{4}\left[\frac{(|\downarrow\rangle_1|\downarrow\rangle_4 + |\uparrow\rangle_1|\uparrow\rangle_4)}{\sqrt{2}}\right][|\varphi^+\rangle_{12}|\psi^+\rangle_{34} + |\psi^+\rangle_{12}|\varphi^+\rangle_{34} + |\psi^-\rangle_{12}|\varphi^-\rangle_{34} - |\varphi^-\rangle_{12}|\psi^-\rangle_{34}]$$

$$+ \frac{1}{4}\left[\frac{(|\downarrow\rangle_1|\downarrow\rangle_4 - |\uparrow\rangle_1|\uparrow\rangle_4)}{\sqrt{2}}\right][|\varphi^-\rangle_{12}|\psi^+\rangle_{34} + |\psi^-\rangle_{12}|\varphi^+\rangle_{34} + |\psi^+\rangle_{12}|\varphi^-\rangle_{34} - |\varphi^+\rangle_{12}|\psi^-\rangle_{34}]$$

$$+ \frac{1}{4}\left[\frac{(|\downarrow\rangle_1|\uparrow\rangle_4 + |\uparrow\rangle_1|\downarrow\rangle_4)}{\sqrt{2}}\right][|\varphi^+\rangle_{12}|\varphi^+\rangle_{34} - |\varphi^-\rangle_{12}|\varphi^-\rangle_{34} + |\psi^+\rangle_{12}|\psi^+\rangle_{34} + |\psi^-\rangle_{12}|\psi^-\rangle_{34}]$$

$$+ \frac{1}{4}\left[\frac{(|\downarrow\rangle_1|\uparrow\rangle_4 - |\uparrow\rangle_1|\downarrow\rangle_4)}{\sqrt{2}}\right][|\varphi^-\rangle_{12}|\varphi^+\rangle_{34} - |\varphi^+\rangle_{12}|\varphi^-\rangle_{34} + |\psi^+\rangle_{12}|\psi^-\rangle_{34} + |\psi^-\rangle_{12}|\psi^+\rangle_{34}]$$

(3)

This formula shows which spin-spin Bell state is selected as a function of the results of the photonic Bell state measurements at Alice's and Bob's locations. For instance, the selection of state $|\varphi^+\rangle_{12}$ and $|\psi^+\rangle_{34}$ after Bob's and Alice's Bell state measurement will project the spin-spin state in the entangled state $(|\downarrow\rangle_1|\downarrow\rangle_4 + |\uparrow\rangle_1|\uparrow\rangle_4)/\sqrt{2}$. All the photons are measured in the user's BSM device leaving the floor to a spin-spin entangled state between Alice's and Bob's network ions.

At this step, the photonic network has successfully established an entanglement between the trapped ions and the quantum circuit can begin: local controlled-Z gates can be performed between the network and the circuit ion qubits, at both Alice's and Bob's side (Figure **3-3**). Afterwards, mid-circuit measurements of the network qubits in the *X* and *Y* bases in Alice and Bob are performed, respectively. Alice and Bob exchange the measurement outcomes in real time and perform single-qubit feed-forward operations ($U_A$, $U_B$) conditioned on the exchanged bits to complete the gate teleportation protocol (see Methods of Ref [5] for the value of $U_A$ and $U_B$ depending on the results of the measurements). This procedure implements the non-local gate $|\psi_{in}^{AB}\rangle \rightarrow U_{CZ}^{AB}|\psi_{in}^{AB}\rangle$, with $|\psi_{in}^{AB}\rangle$ the initial circuit qubits arbitrary state. The number of the teleported controlled-Z gate is therefore proportional to the number of entangled states $|\psi^+\rangle_{23}$ delivered by the space-based QIN, assuming a good enough fidelity.

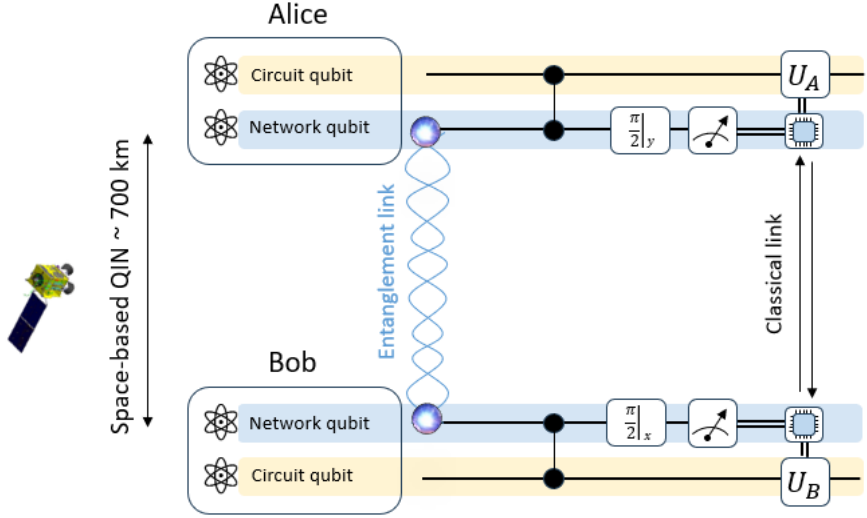

Figure 3-3: Teleportation of a controlled-Z gate between Alice and Bob ions. The entanglement link is provided by the space-based QIN (adaptation of Ref [5]).

Note that the use case under consideration could be extended to blind quantum computing which provides a way for a client to execute a quantum computation using one or more remote quantum servers while keeping the structure of the computation hidden [26].

## 4. SPACE-BASED QIN PERFORMANCES

Our main goal is to give a rough order of the number of controlled-Z gates that could be teleported during the satellite passage over the two quantum ground stations in Calern and in Palaiseau in dual visibility and clear sky (i.e., no clouds). This passage with dual quantum ground station visibility represents roughly $T_{pass} = 331$ seconds in our simulation, with our selected parameters in Table 1 in the Annex.

Before considering the entire chain from Alice to Bob, we start by focusing on the number of entangled photon pairs sent by the satellite and received between Calern and Palaiseau quantum ground stations during the satellite pass. Figure 4-1 (left) shows the rate of entangled photon pairs,

$$\sigma_{elem}^{sat}(Time) = R_{src} \cdot \eta_{elem}^{sat}(Time),  \qquad (4)$$

received at the two quantum ground stations over time during the satellite pass. We observe a maximum of ~4000 entangled photon pairs received at the quantum ground station per second at the maximum satellite elevation, for which the single path link budget (Eq. (13)) has lower losses, see Figure 4-1 (right).

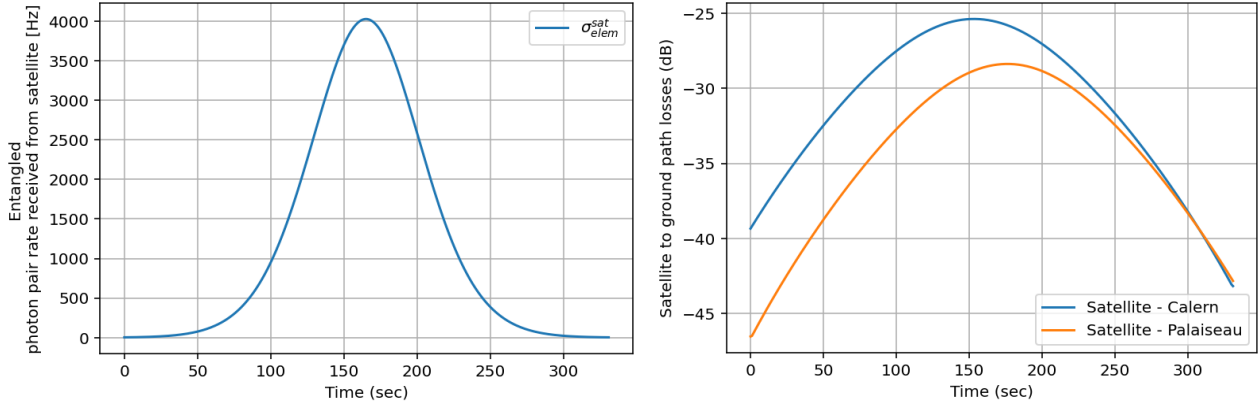

Figure 4-1: (Left) Rate of entangled photon pairs $\sigma_{elem}^{sat}(t)$ (Eq.(4)) received at the MéO station (Calern) and Palaiseau quantum ground stations, distributed over time during the satellite pass. (Right) Single path link budget (Eq.(13)) from the satellite to the quantum ground station.

Figure 4-2 shows the associated cumulated rate,

$$\Sigma_{elem}^{sat}(Time) = \int_0^{Time} \sigma_{elem}^{sat}(t) \cdot dt , \qquad (5)$$

between the two quantum ground stations (blue curve) and between Alice and Bob,

$$\Sigma_{pair}^{end-to-end}(Time) = \int_0^{Time} \sigma_{pair}^{end-to-end}(t) \cdot dt , \qquad (6)$$

at the end of the QIN chain (orange curve), distributed over time during the satellite pass.

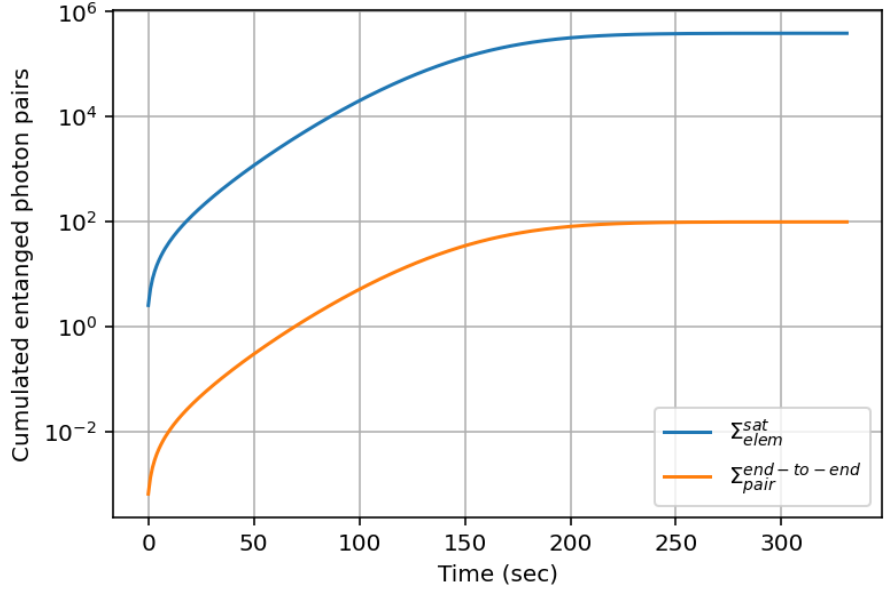

Figure 4-2: (Semi-log scale) Cumulated number of entangled photon pairs, $\Sigma_{elem}^{sat}$ (Eq.(5)), received over time at the MéO and Palaiseau quantum ground stations (blue curve) and integrated entangled photon pairs $\Sigma_{pair}^{end-to-end}$ (Eq.(6)) received between Alice and Bob at the end of the QIN chain (orange curve), distributed over time during the satellite pass.

We observe a saturation of the curves at ~230 seconds at $\Sigma_{elem}^{Sat} \simeq 386\,000\ pairs$ of entangled photon and $\Sigma_{pair}^{end-to-end} \simeq 99$ entangled photon pairs. In other words, Alice and Bob receive only 0.025% of the entangled photon pair received from the satellite at the quantum ground stations. The ground network introduces ~99.975% of losses, among which 93.75% losses are inherent to the four BSM and the rest coming from the optical fiber losses, the quantum memory losses, the frequency conversion losses, etc. This result emphasis the limitation in performance of using many BSM in a chain, although being a strategic advantage for entanglement swapping. In summary, with our system parameters, 99 entangled photon pairs could be used for controlled-Z gate teleportation between Alice and Bob during the satellite passage of 331 seconds.

As explained above, the entangled photon rate is not the only key driver. The signal quality, very often quantified by the fidelity, is a second key performance indicator that should be assessed. Indeed, two users can receive a high-number of entangled photon pairs but with low fidelity, thus making the signal unusable. Figure 4-3 shows the end-to-end fidelity (Eq.(2)) during the satellite pass, considering different straylight levels. We clearly see the negative impact at the beginning and at the end of the satellite pass, when the losses are maximized, see Figure 4-1 (right). Indeed, the signal-to-noise ratio reaches its lowest values and the detector dark count rate introduces noise in the fidelity, leading to the minimal value, i.e. $F \sim 0.25$, for higher straylight level, thus justifying the need of strong straylight filtering. The maximal value of the fidelity is obtained around $t \sim 165\ s$ for higher elevations. The exploitable entangled photon pairs resource for controlled-Z gate teleportation will be available mainly when the satellite is at higher elevations.

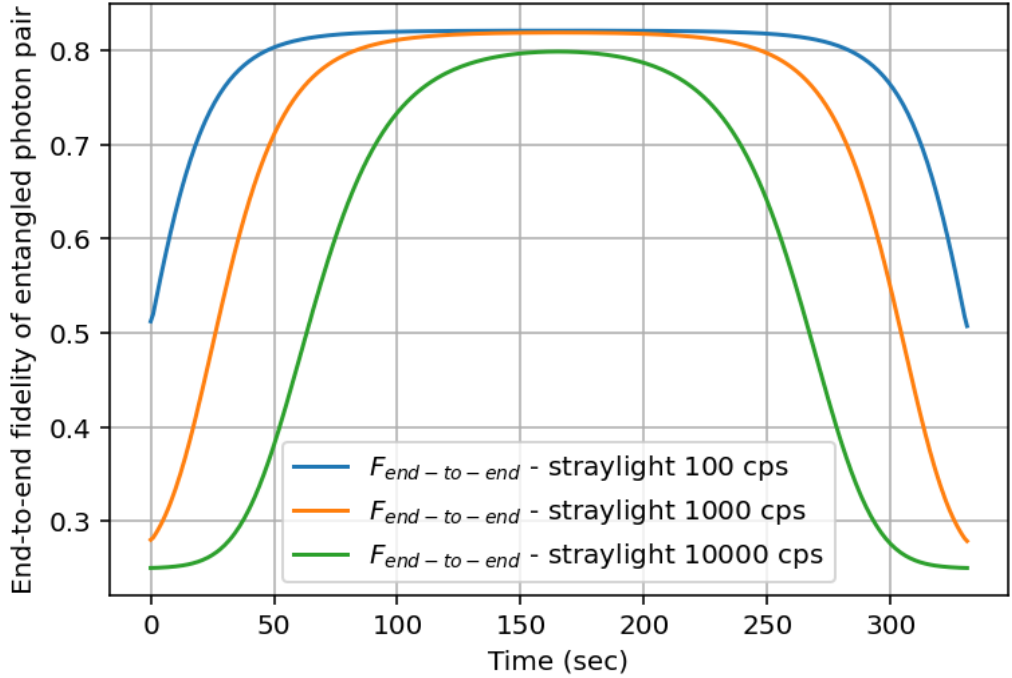

Figure 4-3: End-to-end fidelity Eq.(2) of the received entangled photon pairs over time during the satellite pass, at Alice's and Bob's nodes, for three values of straylight received from space.

## 5. CONCLUSION

This theoretical study emphasizes the role of the satellite in quantum information networks, for the distribution of entanglement towards two remote end-users, Alice and Bob. Particularly, we have focused on the use case of quantum computing where gate teleportation is required. We have extrapolated the spatial range of a state-of-the-art study based on a 2-meter remote trapped-ion qubits setup [5]. We rather consider the range of the France scale for which a fiber-based QIN is highly inefficient, thus requiring the use of a low Earth orbit satellite. We have proposed and assessed theoretically a space-based quantum information network based on existing infrastructures in Paris and Nice. The proposed system allows for the distribution of end-to-end entanglement between Alice in Paris and Bob in Nice, each owing a few qubit processor based on trapped ions. We have developed a model for calculating the rate and fidelity of entanglement distribution to remote end-users through a space-based QIN. We show that, with our system parameters, the system can distribute roughly 90 entangled photon pairs that can be used for controlled-Z gate teleportation between Alice and Bob over the satellite visibility of 331 seconds, reaching the maximal fidelity of 82% for higher elevations of the satellite (during ~100 seconds) and for low straylight level. In this work we considered mostly realistic system parameters although our strong hypotheses on quantum memories have never been demonstrated yet. We have considered a lower performance bond for the entanglement swapping strategy, which could be optimized in a further study to have better end-to-end performances. This study should be considered as a preliminary attempt for proposing a realistic space-based quantum information network at a national scale and gives a rough idea of the service level that could be reached with the considered system parameters. Such a system clearly requires the maturation of quantum memories and entanglement swappers. Our intent is to show that a path exists towards a global Quantum Information Network and perhaps to spur efforts to reach the quantum memory specifications considered here.

## ACKNOWLEDGEMENTS

This work was supported by Thales Alenia Space internal funding. We thank the French Space Agency (Centre National d'Etudes Spatiales - CNES) and the European Space Agency (ESA) for their support on related projects such as QINSAT and TeQuants. We thank the groups of Eleni Diamanti and Sébastien Tanzilli for their advice.

## ANNEX

In this annex we describe our models used for the calculation of the ground and space path performances in terms of entangled photon rate per second on the elementary link and entanglement fidelity. A table of the parameters used for our simulations is also provided.

**Model for the efficiency $\eta_{elem}^{Paris}, \eta_{elem}^{Nice}$ of the ground elementary links**:

The advantage of a chain of quantum repeaters lies in the presence of heralded quantum memories within the nodes. These quantum memories make it possible to reduce the impact of losses in propagation channels (optical fibers for ground networks and free space for satellite links). Indeed, in the context of a simple chain as illustrated in Figure 5-1, if a photon is lost between a source and a node at a given timeslot but at this same instant another pair is received from another source, then the latter can be stored in a memory while a photon from another pair is received.

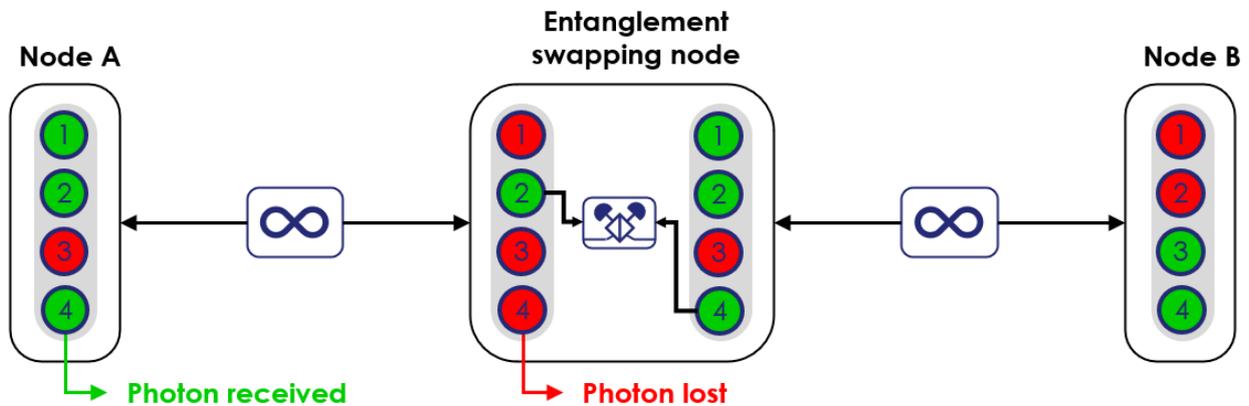

Figure 5-1: Bell state measurement device assisted with quantum memories. The entanglement swapping will be performed by using the pairs #2 and #4.

Figure 5-1 shows nodes with multiplexed quantum memories. That is to say, they have several slots in which qubits arriving at different times can be stored. Nodes A and B of this elementary link have one memory with four slots, and the entanglement swapping (BSM) node has two memories with four slots each (or one memory with eight slots). The memories are indexed with timeslots, meaning that a photon may be written in the first memory slot during the first timeslot, etc.

A timeslot is defined as $\delta t = 1/R_{src}$, with $R_{src}$ the entangled photon pairs source pump rate, and the duration of the time window is defined as $t_k = k \cdot \delta t$ with $k \in [1, N]$ and $N$ the number of memory storage modes.

We see in Figure 5-1 that the entanglement distribution between node A (resp. B) and the entanglement swapping node only worked for timeslot $k = 2$ (resp. $k = 4$). Entanglement swapping is thus possible between the qubits received at timeslots $k = 2$ (resp. $k = 4$), something that would not have been possible without quantum memories. This process is possible with the hypothesis that quantum memories have written announcement (heralding) for each storage mode.

In the example of Figure 5-1, we studied the behavior of a small chain for four timeslots. To do this, it was necessary to take into account quantum memories with four storage slots. In order to generalize this example for $N$ temporal intervals we will consider quantum memories with $N$ storage slots. Each of these slots have its own performance « $\eta_{QM}(t_k)$ », i.e., a probability of successfully writing and then reading the quantum information placed there. Given that we study the system after temporal instants, the performance of the locations will therefore vary (Figure 5-2). During a timeslot, the memory efficiency is considered constant.

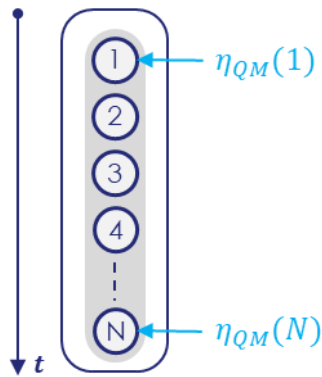

Figure 5-2: Performance of quantum memory locations

The probability of having correctly written and read the information stored in location #1 at the moment $N$ is lower than that of location #$N$. This is explained by the losses of coherence that are at play within a quantum memory: the longer information is stored, the lower the probability of recovering it. The efficiency of quantum memory locations is modeled to follow an exponentially decreasing trend over storage time of the type:

$$\eta_{QM}(t_k) = \eta_{QM,W} \cdot \exp\left(-\frac{t_k}{\tau_{QM}}\right) \qquad (7)$$

where $\eta_{QM,W}$ is the writing efficiency of the memory and $\tau_{QM}$ is its characteristic storage time. We can then apply this formula for the different memory locations according to the desired temporal granularity. $\tau_{QM}$ is expressed as a number of timeslots as $\tau_{QM} = \tau_{QM,sec}/\delta t$ where $\tau_{QM,sec}$ is the characteristic storage time expressed in seconds, and $\delta t$ the duration of one timeslot.

We assume that a chain of quantum repeaters can be divided into multiple elementary links. An elementary link corresponds to the distribution of entanglement from a source to two nodes, as illustrated in Figure 5-3.

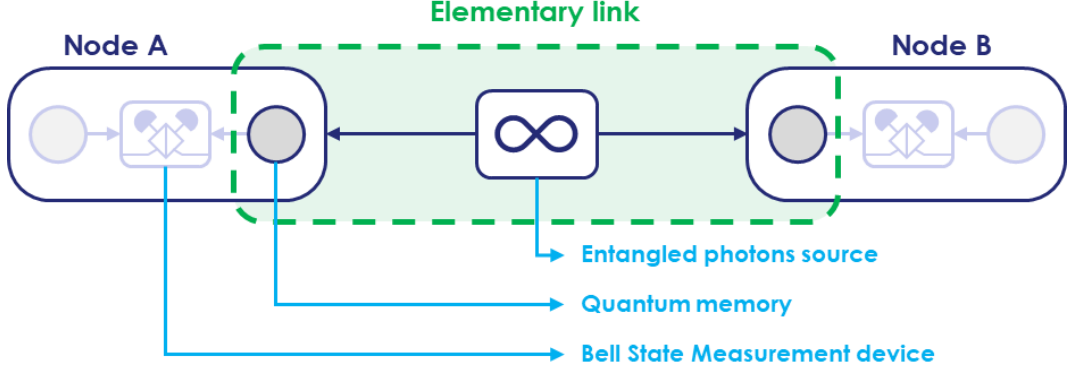

Figure 5-3: Elementary link of a chain of entanglement switches

The objective is to determine the probability of successfully transmitting at least one pair of entangled photons between the two nodes after $N$ timeslots. This objective corresponds to a worst-case scenario for system performance because if two pairs of entangled photons (or more) are received, only one will be kept to carry out the entanglement swapping. This scenario remains interesting for studying the optimal number of temporal instants after which to look at the state of quantum memories, and therefore optimize their use.

An elementary link is made up of three types of elements:
- A source of entangled photon pairs;
- Two propagation channels (on either side of the source);
- Two quantum memories.

The entangled photon source is characterized by an entangled photon pair generation efficiency $\eta_{src}$, and a source rate $R_{src}$.

The propagation channels are characterized by their length $l$ and their efficiency $\eta_{ch}$ which correspond to the efficiency of reception of a photon coming from the entangled photon source. This efficiency depends on the channel length for fiber optic propagation (assuming 0.2 dB/km losses).

Quantum memories are characterized by the efficiency $\eta_{QM}(t_k)$ of their different modes, where $k \in [1, N]$ with $N$ the number of memory storage modes. $\eta_{QM}(t_k)$ is a function of additional parameters such as the writing probability $\eta_{QM,W}$ and the characteristic time $\tau_{QM}$.

The probability of failing to transmit the entanglement between two nodes at timeslot $t_k$ is:

$$p_{fail}(t_k) = 1 - \eta_{src} \cdot \eta_{ch1} \cdot \eta_{ch2} \cdot \eta_{QM}^2(t_k) \tag{8}$$

The probability of failing to transmit the entanglement between two nodes after $N$ time instants is:

$$p_{all\ fail} = \prod_{t_k=1}^{N} \left(1 - \eta_{src} \cdot \eta_{ch1} \cdot \eta_{ch2} \cdot \eta_{QM}^2(t_k)\right) \tag{9}$$

Thus, the efficiency of the ground elementary link is given by the probability of transmitting at least one pair of entangled photons on an elementary link after $N$ timeslots is:

$$\eta_{elem} = 1 - \prod_{t=1}^{N} \left(1 - \eta_{src} \cdot \eta_{ch1} \cdot \eta_{ch2} \cdot \eta_{QM}^2(t)\right) \tag{10}$$

$\eta_{elem}^{Paris}, \eta_{elem}^{Nice}$ are obtained by considering the ground link efficiency $\eta_{ch} = 10^{\frac{-0.2 \cdot l}{10}}$ for distance $l$ in kilometers and considering 0.2 dB losses per kilometer in the optical fiber. We take $l = (14\ km, 45\ km)$ for Paris and $l = (32\ km, 35\ km)$ for Nice, as shown in Figure 3-1. $\eta_{elem}$ can also be used for the space link by using the space link budget Eq.(13).

A chain of entanglement switches is only a succession of elementary links, to which Bell state measurement modules are added (Figure 5-4).

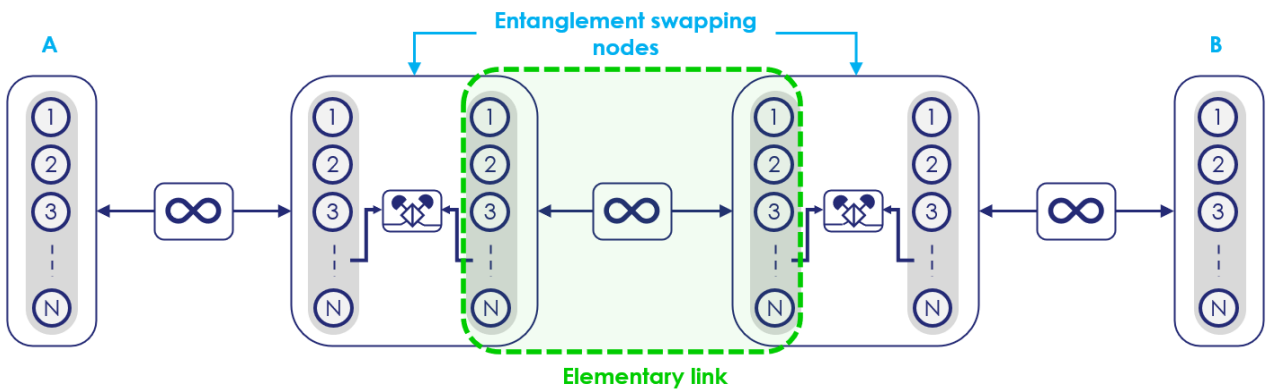

Figure 5-4: Quantum repeaters chain with elementary link

Figure 5-4 shows a complete ground chain, with two end users A and B, as well as two quantum repeaters. A quantum repeater is defined as a Bell State Measurement module (BSM) and single photon detectors. For two repeater nodes we therefore have: 3 elementary links and 2 entanglement swapping nodes. Two new components appear in the performances of a repeater chain:
- Bell State Measurement device (BSM);
- Single photon detectors.

The BSM is characterized by an efficiency $\eta_{BSM}$ related to the projection of the quantum state onto a useful Bell state (this efficiency is necessarily smaller than ½ since only two of the Bell states can be identified with linear optics). Single photon detectors are characterized by an efficiency $\eta_{det}$. A BSM requires two detectors simultaneously (four are required in total, but only two detectors are used per BSM during a Bell state measurement).

**Model for the efficiency $\eta_{elem}^{Sat}$ of the space elementary link:**

The space path is composed of one satellite, two ground stations and two quantum memories (Figure 5-5). The efficiency of the space link is given by:

$$\eta_{elem}^{Sat} = \eta_{src} \cdot \eta_A \cdot \eta_B \cdot \eta_{QM}^2(t_k), \quad (11)$$

with

- $\eta_{src}$ the efficiency of the entangled photons source, see Table 1;
- $\eta_A$ the channel efficiency of channel A, see Eq. (13);
- $\eta_B$ the channel efficiency of channel B, see Eq. (13);
- $\eta_{QM}(t_k)$ the efficiency of the quantum memory, see Eq. (7).

The channel efficiencies are given by the space-to-ground link budget described hereafter. The received power of an optical signal sent from a satellite and received by a telescope on the ground is given by:

$$P_R = P_0 \cdot G_T \cdot \eta_{atm} \cdot G_R \quad (12)$$

with:

- $P_0$ the power of the entangled photon source
- $G_T = \eta_{cT}\left(1 - e^{-\frac{D_T^2}{2\omega_0^2}}\right)$ the transmitter gain (telescope efficiency). $\eta_{cT}$ is the internal transmittance of the on-board telescope (including wave front errors and obscuration), $D_T$ is the telescope aperture, and $\omega_0 = \frac{D_T}{2}$ is the beam waist at the output of the telescope.
- $\eta_{atm}$ the downlink atmospheric losses (absorption, scattering, cirrus clouds, aerosols). $\eta_{atm} = (\eta_{atm,0})^{\frac{1}{\sin(\theta)}}$ where $\theta$ is the satellite elevation and $\eta_{atm,0}$ the atmospheric losses when the satellite is at zenith ($\theta = \pi/2$).
- $G_R = \eta_{cR}\left(1 - e^{-\frac{D_R^2}{2\omega(r)^2}}\right)$ the receiver gain at the telescope level (telescope efficiency). $\eta_{cR}$ is the internal losses of the receiver telescope (including wave front errors, obscuration, optical fiber coupling with adaptive optics), $D_R$ is the telescope aperture, and $\omega(r)$ the received beam waist at the ground station side (with $\omega_0 = D_{sat}$), with $r$ the free space length between the satellite and the receiver's telescope pupil.

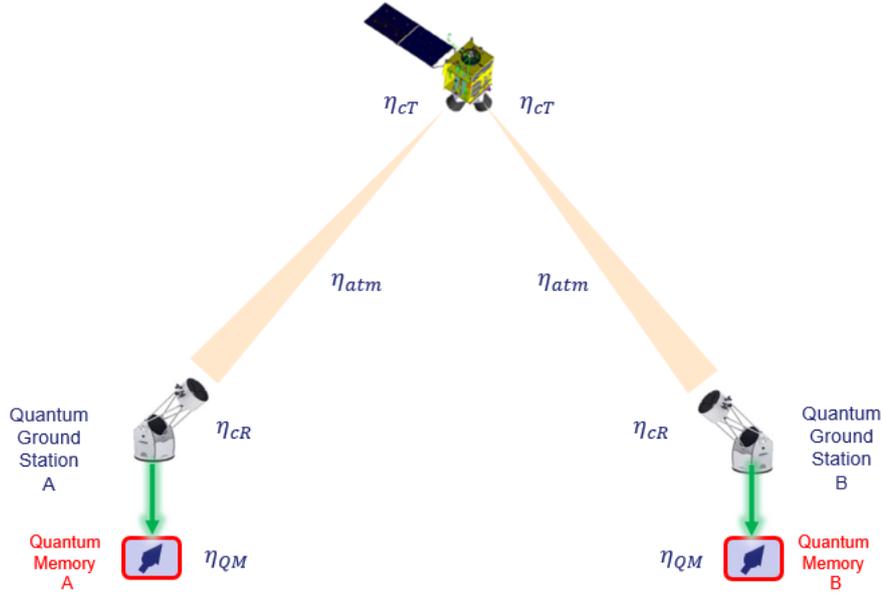

Figure 5-5: Space path connecting two quantum memories, each connected to a quantum ground station. The parameters $\eta$ represent the efficiency on each part of the channel (see text).

Assuming Gaussian optics, the beam waist is given by $\omega(r) = \omega_0\sqrt{1 + \left(\frac{r}{R_R}\right)^2}$, with the Rayleigh length $R_R = \frac{\pi\omega_0^2}{\lambda}$ ($\lambda$ is the signal wavelength).

Finally, the single path link budget is given by:

$$\eta_{A,B} = \frac{P_{R,A,B}}{P_0} = \eta_{cT} \cdot (1 - e^{-2}) \cdot \eta_{atm} \cdot \eta_{cR} \left[1 - e^{-\frac{2D_R^2}{D_T^2}\left(\frac{1}{1+\frac{16\lambda^2 r_{A,B}^2}{\pi^2 D_T^4}}\right)}\right] \quad (13)$$

**Model for the Werner parameter of the end-to-end path $\mathcal{W}_{end-to-end}$:**

The fidelity of the propagated entangled photon pair is calculated with the following model, based on Werner states [27]. Similarly to the previous case on throughput modeling, we can model the evolution of fidelity by sections of elementary links. In order to evaluate the impact of unwanted photons on the overall fidelity of distributed states on the network, a decoherence channel model is assumed. For example, the states produced by the source are considered to be Werner states (noisy EPR pairs) [27] with an associated density matrix of the following form:

$$\rho_{src} = \mathcal{W}_{src}|\psi^+\rangle\langle\psi^+| + \frac{1-\mathcal{W}_{src}}{4}\mathbb{I}_4 \quad (14)$$

where $|\psi^+\rangle$ is a perfectly entangled Bell state, $\mathcal{W}_{src}$ is the Werner parameter of the source associated with its fidelity $F_{src}$ with respect to $|\psi^+\rangle$, and $\mathbb{I}_4$ is a 4×4 identity matrix.

The Werner parameter of an elementary link can be estimated by concatenating the Werner parameters of each functional block. The source of entangled photons is characterized by $F_{src}$ fidelity because the latter may not produce perfect Bell states. We consider $\mathcal{W}_{src} = F_{src}$.

The noisy propagation channel is characterized by a fidelity $F_{fiber}$. Indeed, optical fibers induce polarization and phase rotations, which are complicated to characterize and deviations in the measurement bases can be caused, which will also impact the fidelity of the qubits. We consider $\mathcal{W}_{fiber} = F_{fiber}$.

Quantum memories are characterized by a $F_{QM}$ fidelity. When quantum memory re-emits a photon, its quantum state has interacted with the memory and is no longer exactly the same as when it was written. We consider $\mathcal{W}_{QM} = F_{QM}$.

We can model the Werner state of a ground elementary link $\mathcal{W}_{elem}$ as:

$$\mathcal{W}_{elem} = \mathcal{W}_{src} \cdot \mathcal{W}_{fiber}^2 \cdot \mathcal{W}_{QM}^2 \quad (15)$$

The extension of this formula to a chain must take into account the Bell state measurement devices (BSM) including the single photon detectors. We will assume that only single photon detectors inside the BSM can have an impact on fidelity degradation, e.g. due to the dark counts or stray photons. Single photon detectors are characterized by a "dark count" rate $R_{dc}$. The latter designates detections by a photodetector in the absence of incident light. They are caused by sources of noise intrinsic to the detector itself, such as thermal or electronic noise. A dark count detection can blind the detection of the photon of interest, which would induce a measurement error. To establish the effect of these dark counts on the fidelity, it is necessary to take into account the generation rate of the sources of entangled photons $R_{src}$ and the efficiency of the elementary links $\eta_{elem}$ (Eq. (10)).

A click on a detector can come from two sources with different rates:
- Click rate of detection of photon coming from the source $R_{true} = R_{src} \cdot \eta_{elem}$;
- False click rate $R_{false} = R_{dc}$ (due to dark counts, etc. photon not coming from the source).

$R_{true}$ is the rate at which pairs of photons from the source, after attenuation in the channel and storage in memories, reach the detectors and are detected. This rate does not take into account pairs of photons for which one of the two photons would have been lost/absorbed.

The detector can either click because of a real photon measurement or because of a dark count measurement. The total click-through rate is therefore the sum of the two rates:

$$R_{total} = R_{true} + R_{false} = R_{src} \cdot \eta_{elem} + R_{dc} \quad (16)$$

The probability of having a click of a photon coming from the source is given by:

$$P(true \,|click) = \frac{R_{true}}{R_{total}} = \frac{R_{src} \cdot \eta_{elem}}{R_{src} \cdot \eta_{elem} + R_{dc}} \quad (17)$$

Eq.(17) can be used to obtain the Werner parameter for two connected elementary ground links labeled $i$ and $i + 1$:

$$\mathcal{W}_{BSM,i} = \frac{R_{src} \cdot \eta_{elem,i}}{R_{src} \cdot \eta_{elem,i} + R_{dc}} \times \frac{R_{src} \cdot \eta_{elem,i+1}}{R_{src} \cdot \eta_{elem,i+1} + R_{dc}} \tag{18}$$

Eq.(18) can be easily extrapolated to the terrestrial chain of nodes composed of $M$ elementary links and $M - 1$ swapping nodes:

$$\mathcal{W}_{end-to-end} = \prod_{i=1}^{M} \mathcal{W}_{elem,i} \cdot \prod_{i=1}^{M-1} \mathcal{W}_{BSM,i} \tag{19}$$

In the architecture under consideration (Figure 3-1), this leads to:

$$\mathcal{W}_{end-to-end} = \mathcal{W}_{src}^{Alice} \cdot \mathcal{W}_{BSM}^{Alice} \cdot \mathcal{W}_{elem}^{Paris} \cdot \mathcal{W}_{BSM}^{TRT} \cdot \mathcal{W}_{elem}^{Sat} \cdot \mathcal{W}_{BSM}^{Calern} \cdot \mathcal{W}_{elem}^{Nice} \cdot \mathcal{W}_{BSM}^{Bob} \cdot \mathcal{W}_{src}^{Bob} \tag{20}$$

with $\mathcal{W}_{elem}^{Paris,\ Nice}$ given by Eq. (15) and $\mathcal{W}_{src}^{Alice} = \mathcal{W}_{src}^{Bob} = F_{src} \cdot F_{conv}$.

During a free-space link from a satellite, stray photons from the environment can be received in the telescopes and can be coupled to the single mode fiber at ground stations. At the detection level, these photons will have an impact on the fidelity of the quantum states, as they may be confused with useful photons sent by the satellite's payload. The rate of stray photons is strongly dependent on the acquisition conditions, with or without the presence of the Moon for nighttime operations and considering also the possible Sun satellite reflection. The effect of these stray photons on the system's fidelity can be introduced directly into the detector noise at Calern and TRT sites (i.e. in $\mathcal{W}_{BSM}^{BSM}$ and $\mathcal{W}_{BSM}^{TRT}$), in addition of dark counts. The space path is composed of a single elementary link with an entangled photon source, two propagation free space channels and two quantum memories. Similarly to the ground path, the Werner parameter of this elementary link $\mathcal{W}_{elem}^{Sat}$ can be computed from Eq.(15) using $F_{freespace}$ instead of $F_{fiber}$.

**Satellite Orbit:**

We have used ANSYS Systems Tool Kit® for the generation of the orbital simulations for a satellite at 600 km altitude, with 60-degree inclination and right ascension of the ascending node (RAAN) of 72.5 degrees, on date 01/01/2025 – 00:00:00, during nighttime. The satellite's ground trace and the station positions are plotted in Figure 5-6. The dual communication time for the entanglement distribution is 331 seconds. We do not consider clouds.

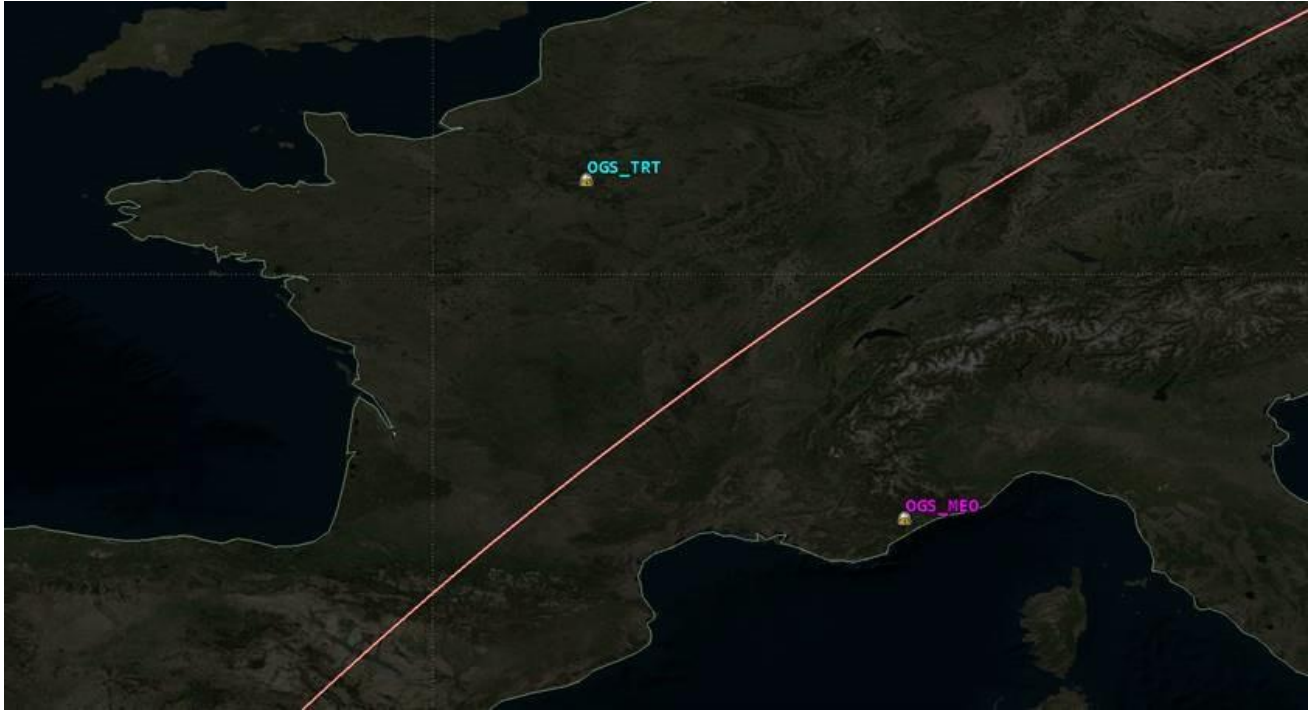

Figure 5-6: Satellite's ground trace over France

**List of parameters**:

Table 1 details the various parameters taken into account in our simulations.

**Table 1 : Parameters list used in our simulations**

| Parameter | Symbol | Value |
|---|---|---|
| Satellite altitude | | $600\ km$ |
| Satellite inclination | | 60 degrees |
| Right ascension of the ascending node (RAAN) | | 72.5 degree |
| Start simulation date | | 01/01/2025 – 00:00:00 nighttime |
| Entangled photon source wavelength | $\lambda$ | $1550\ nm$ |
| Entangled photon source efficiency | $\eta_{src}$ | 0.25 |
| Entangled photon source rate | $R_{src}$ | $1\ GHz$ |
| Wavelength conversion efficiency | $\eta_{conv}$ | 0.8 |
| Wavelength conversion fidelity | $F_{conv}$ | 0.98 |
| Entangled photon source fidelity | $F_{src}$ | 0.99 |
| Quantum memory writing efficiency | $\eta_{QM}$ | 0.98 |

| Parameter | Symbol | Value |
|---|---|---|
| Quantum memory fidelity | $F_{QM}$ | 0.98 |
| Quantum memory storage modes | $N$ | 500 |
| Quantum memory characteristic storage time | $\tau_{QM}$ | $10\ ms$ |
| Quantum memory storage window | $\omega_{storage}$ | $250\ ps$ |
| SNSPD detector efficiency | $\eta_{det}$ | 0.9 |
| SNSPD detector dark count rate for the ground network | $R_{dc}$ | $50\ cps$ |
| BSM efficiency | $\eta_{BSM}$ | 0.5 |
| Optical fiber attenuation | $\alpha$ | $0.2\ dB/km$ |
| Optical fiber link fidelity | $F_{fiber}$ | 0.99 |
| Tx telescope diameter | $D_{TX}$ | $40\ cm$ |
| Rx telescope diameter MéO station @Calern | $D_{RX}$ | $150\ cm$ |
| Rx telescope diameter station @THALES Palaiseau | $D_{RX}$ | $100\ cm$ |
| Free-space link fidelity | $F_{freespace}$ | 0.99 |
| Atmospheric transmittance when the satellite is at zenith | $\eta_{atm,0}$ | 0.2 |
| Internal transmittance of the ground telescope (adaptive optics included for single mode fiber coupling) | $\eta_{cR}$ | 0.1 |
| Internal transmittance of the on-board telescope | $\eta_{cT}$ | 0.7 |